\documentclass{llncs}
\usepackage{mathptmx}       
\usepackage{helvet}         
\usepackage{courier}        
\usepackage{type1cm}        
\usepackage{makeidx}         
\usepackage{graphicx}        
\usepackage{multicol}        
\usepackage[bottom]{footmisc}
\usepackage{subfigure}
\usepackage{amsfonts}
\usepackage[cmex10]{amsmath}
\usepackage{fixltx2e}
\usepackage{amsmath}
 \begin{document}

\title{Adaptive rewiring evolves brain-like structure in directed networks}
\titlerunning{Adaptive rewiring in brain}  
%
\author{Ilias Rentzeperis\inst{1} \and Steeve Laquitaine\inst{1}  \and Cees van Leeuwen\inst{1, 2}}
\authorrunning{Ilias Rentzeperis et al.} 
%
\tocauthor{Ilias Rentzeperis, Steeve Laquitaine, Cees van Leeuwen}
\institute{KU Leuven, Belgium,\\
	\email{cees.vanleeuwen@kuleuven.be},\\ 
	\and
	University of Technology Kaiserslautern, Germany}

\maketitle    

\abstract{Brain networks are adaptively rewired continually, adjusting their topology to bring about functionality and efficiency in sensory, motor and cognitive tasks.  In model neural network architectures, adaptive rewiring generates complex, brain-like topologies. Present models, however, cannot account for the emergence of complex directed connectivity structures. We tested a biologically plausible model of adaptive rewiring in directed networks, based on two algorithms widely used in distributed computing: advection and consensus. When both are used in combination as rewiring criteria, adaptive rewiring shortens path length and enhances connectivity. When keeping a balance between advection and consensus, adaptive rewiring produces convergent-divergent units consisting of convergent hub nodes, which collect inputs from pools of sparsely connected, or local, nodes and project them via densely interconnected processing nodes onto divergent hubs that broadcast output back to the local pools. Convergent-divergent units operate within and between sensory, motor, and cognitive brain regions as their connective core, mediating context-sensitivity to local network units. By showing how these structures emerge spontaneously in directed networks models, adaptive rewiring offers self-organization as a principle for efficient information propagation and integration in the brain.}

\section{Introduction}
Neuronal selectivity not only depends on local receptive fields, but also on global contextual features \cite{Alexander}. Prominent examples are surround suppression \cite{Hubel}\cite{Blakemore}, sensorimotor prediction coding \cite{Keller}, orientation biases \cite{Sasaki}, and normalization \cite{Carandini}. These and other contextualized effects are mediated by short- and long-range connections within areas \cite{Das}\cite{Keller2}, and top-down feedback \cite{Hupe}. A common mechanism accommodating contextual effects is pooling, i.e. units collect inputs from many neurons which and then redistribute the information back to the neurons \cite{Carandini2}. This process can be accomplished via convergent-divergent units \cite{Kumar}. Convergent-divergent units receive input from many neurons (via convergent connections), process it, and subsequently project the output to many neurons (via divergent connections).   
\newline

\par We investigated how a network connectivity structure encompassing convergent-divergent units could emerge spontaneously from adaptive plasticity in artificial neural networks.  It is well known that the brain’s connectivity structure continually evolves during development \cite{Sur}, learning \cite{Plautz} and recovery following injury \cite{Nudo}\cite{Nudo2}. Whereas most studies of adaptive plasticity have been focused on activity evoked by external (i.e. sensory) input \cite{Hardy}\cite{Chrol}, a major part of brain activity  is ongoing and spontaneous \cite{Sadaghiani}. To investigate whether convergent-divergent units can emerge from spontaneous activity, we used a simple, generic structural plasticity \cite{Butz} mechanism characterized as adaptive rewiring: strong activity between neural components leads to the addition of network connections, while weak activity leads to their pruning.
\newline

\par Previous studies on adaptive rewiring have shown that spontaneous activity can lead to complex network structures, akin to the anatomy of the brain: modular or centralized networks with rich club effect \cite{Rentzeperis}-\cite{Berg}. However, these studies typically used undirected networks. Undirected networks propagate activity indiscriminately in both directions of an edge connecting two nodes. While this property is mathematically convenient and can capture some of the aggregate effects of adaptive rewiring, it is physiologically implausible at the neuronal level, in particular because of the asymmetry in signal transfer through chemical synapses \cite{Sheng}-\cite{Harlow}. Moreover, undirected networks cannot represent brain features that show directionality, such as convergent-divergent units. Directed networks, therefore, offer a more plausible framework for adaptive rewiring. In addition, such a framework is more general than undirected networks, as the latter are included as a special case. 
\newline

\par A recent study proposed to represent neuronal signal propagation as random walks on an anatomical network, and characterize the propagation stochastically in terms of heat diffusion \cite{Abdelnour}. We previously adopted a similar principle in an adaptive rewiring model with undirected connections, where activity propagation between nodes was represented by a heat kernel \cite{Rentzeperis}\cite{Jarman2}. In such models, adaptive rewiring amounts to adding shortcuts to regions with intense diffusion, while pruning where diffusion is low, thereby relating the connectivity structure of the network to neuronal traffic.
\newline

\par To model adaptive rewiring in directed networks, we use two algorithms closely related to heat diffusion that can account for directionality of the connections: advection and consensus \cite{Chapman}. Advection and consensus algorithms have been used extensively in sparsely connected neural networks. In the context of neuronal dynamics, both advection and consensus (and diffusion in the undirected case) act as homeostatic factors that aim to reduce the differences between neurons’ activities. Neuronal activity values are interpreted as concentrations, and traffic between them is proportional to their concentration differences. The goal of advection and consensus algorithms is to approach a common concentration value for all the nodes \cite{Chapman}-\cite{Ren}. 
\newline

\par For both advection and consensus, we applied the same rewiring principle: during each rewiring iteration, a connection with low traffic is cut and used for an unconnected pair of nodes with high activity. We found that when a random network rewires based on advection, the emerging network contains converging hub nodes, which receive pooled inputs from many local (sparsely connected) nodes. When rewiring is based on consensus, the network develops diverging hub nodes, which broadcast outputs back to the local nodes. When both advection and consensus are applied within the same rewiring scheme, the network produces both diverging and converging hubs nodes. Intermediate nodes, nodes on the path between a converging and diverging hub nodes, are densely interconnected, multiple times more densely than the local nodes outside the unit. In addition, a proportion of random rewiring facilitates the formation of these structures and improves node-to-node communication. 
\newline

\par Converging and diverging hub nodes, together with a dense network of intermediate nodes, constitute converging-diverging units. These structures arise in our model in a self-organized fashion, as the product of adaptive rewiring, while the remaining network is sparse. We conclude that adaptive rewiring offers a principle for the emergence of efficient context-sensitive information processing in the brain.

\section{Methods}
\subsection{Digraph preliminaries}
A directed binary graph (or digraph) is defined as an ordered pair, $D = (N, E)$, where $N$ corresponds to the set of nodes, $N = [1, … , n]$ and $E$ to ordered node pairs denoting the directed edges (or connections), $E\subset\ N\times\ N$. The ordered pair $(x,y)\in E$ if x and y are adjacent so that x is the tail of the connection and y its head, also denoted as $x\ \rightarrow y$; if not, then $\left(x,y\right)\notin E$. The cardinalities $\left|V\right|$ and $\left|E\right|$ are the total number of nodes and edges of D respectively.  For a digraph with n vertices, the connectivity pattern is encapsulated by an $N\times\ N$ adjacency matrix A. For binary digraphs all connections carry the same weight, i.e. $A_{ij}=1$ indicates that $(j,i)\in E$, while $A_{ij}=0$ that $\left(j,i\right)\notin E$. We do not allow for a node to point to itself, so A is zero along its diagonal.
\newline
\par The in-degree of node k is the number of connections that have node k as their head; the tails of these connections constitute the in-degree neighborhood of $k$, $N_{in}\left(k\right)$. The number of nonzero elements along the rows of $A$ indicate the in-degrees of the nodes (Fig 1, row of A). The out-degree of node $k$ is the number of connections that have node $k$ as their tail with the corresponding heads being the out-degree neighborhood of $k$, $N_{out}\left(k\right)$. The numbers of nonzero elements along the columns of $A$ indicate the out-degrees of the nodes (Fig 1, column of A).  

\begin{figure}
\centering
\includegraphics[]{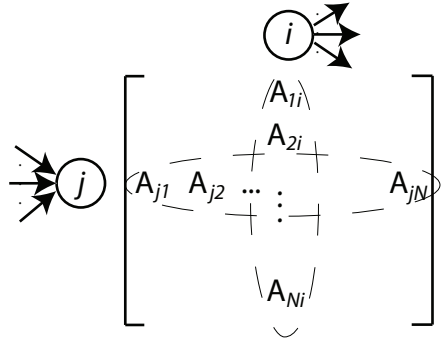}
\caption{Schematic representation of the adjacency matrix
From the rows of A, we obtain the in-degree connectivity of the network, from the columns the out-degree.}
\label{a1:fig:Res}       
\vspace{-0.4cm}
\end{figure}

\subsection{Advection and consensus dynamics }

We use advection and consensus algorithms based on the heat (or diffusion) equation. The heat equation describes how some quantity diffuses across a medium over time. When operating on an undirected graph, the heat equation has an explicit solution: 

\begin{equation}
\label{Eq:equ11}
\\\mathbf{x}\left(t\right)=\ e^{-Lt}\mathbf{x}\left(0\right)\\ 
\end{equation}

where $\mathbf{x}\left(0\right)$ and $\mathbf{x}\left(t\right)$ are n-dimensional vectors containing the concentrations of the nodes of the network, $x_1...x_n$ before and at time t after diffusion respectively. $L$, the graph Laplacian, is defined as $L = D-A$, where $D$ is a diagonal matrix with the degrees of the nodes in its diagonal entries and $A$ the adjacency matrix. It is the discrete analogue of the Laplace-Beltrami operator $(\nabla^2f)$ for graphs. In its discrete form, it emerges in optimization problems such as graph partitioning \cite{Jianbo}, and dimensionality reduction \cite{Belkin}. In the diffusion context, the graph Laplacian is an $N\times\ N$ matrix that quantifies the rate of change of flow across edges. 
\newline  
\par  Mathematically, the difference between advection and consensus is their use of different Laplacian matrices: the former uses the out-degrees of the nodes in the diagonal entries of the Laplacian the latter the in-degrees. Both the advection and consensus dynamics will produce the same values as heat diffusion if the graph edges are stripped of their directionality.  
\newline
\par Below, we give a brief description of the advection and consensus dynamics. A more detailed description is given elsewhere\cite{Chapman}. The advection equation in a continuous manifold is given by: 

\begin{equation}
\label{Eq:equ2}
\frac{\partial u}{\partial t}=\ -\nabla\bullet\left(\vec{y}u\right)
\end{equation}

where $\nabla$ is the divergence operator, and $\vec{y}$ a vector field that affects how the concentration $u$ of a material is distributed. Advection, for example, can describe how the concentration of oil or other pollutants changes within a river stream. 
\newline

\par  In a digraph setting, the direction and weight of the edges define the vector field $\vec{y}$, and the nodes’ values the concentration $u$. The rate of change of the concentration of node i is defined as the flow into the node minus the flow out of the node:   

\begin{equation}
\label{Eq:equ3}
{\dot{x}}_i\left(t\right)=\sum_{{\forall\ j|j\rightarrow i}}{w_{ij}x_j\left(t\right)}-\sum_{\left\{\forall\ k\middle|\ i\rightarrow k\right\}}{w_{ki}x_i\left(t\right)}
\end{equation} 

where at time $t$, $w_{ij}x_j(t)$ indicates the flow through the edge  $j\rightarrow i$. Collectively, the advection dynamics can be represented in matrix form as:  

\begin{equation}
\label{Eq:equ4}
\dot{\mathbf{x}}\left(t\right)=\ -\mathcal{L}_{out}\mathbf{x}\left(t\right)
\end{equation}   

where at time $t$,  $\dot{\mathbf{x}}\left(t\right)=\left[{\dot{x}}_1\left(t\right),\ldots,\ {\dot{x}}_n\left(t\right)\right]^T$ contains the rate of change of the concentration of each node, $\mathbf{x}\left(t\right)=\left[{x}_1\left(t\right),\ldots,\ {x}_n\left(t\right)\right]^T$ the concentration of the nodes, and $\mathcal{L}_{out}$, the out-degree Laplacian, an $N\times\ N$ matrix with the following entries:

\begin{equation*}
\mathcal{L}_{out} = \begin{cases}
\sum_{k=1}^{n}A_{ki} &\text{if $i=j$}\\
-A_{ij} &\text{if $i\neq j$}
\end{cases}
\end{equation*}

The out-degree strengths of the nodes are in the diagonal entries of $\mathcal{L}_{out}$. The solution to (4) is:
 \begin{equation}
\label{Eq:equ5}
 \mathbf{x}\left(t\right)=\ e^{{-L}_{out}t}\mathbf{x}\left(0\right)
 \end{equation}   

For advection, the concentrations of the nodes are sum conservative, i.e.  for any time $t$ $\sum_{i=1}^{n}{x_i\left(t\right) = \sum_{i=1}^{n}{x_i\left(0\right)\ }\ }$. The consensus dynamics for each node is given by:

 \begin{equation}
\label{Eq:equ6} 
 \dot{x_i}\left(t\right)=\sum_{{\forall\ j|j\rightarrow i}}{w_{ij}(x_j\left(t\right)-x_i\left(t\right))}\ \ \ \ i=1,\ldots,n 
 \end{equation}   
 
Eq. (6) implies that there is diffusion whenever there is a difference in concentrations between adjacent nodes.

\par Collectively, the consensus dynamics (6) are represented in matrix form as: 
 \begin{equation}
\label{Eq:equ7} 
\dot{\mathbf{x}}\left(t\right)=\ -\mathcal{L}_{in}\mathbf{x}\left(t\right)
 \end{equation}   

where $\mathcal{L}_{in}$ is the in-degree Laplacian, an $N\times\ N$ matrix with the following entries:
 
 \begin{equation*}
\mathcal{L}_{in} = \begin{cases}
\sum_{k=1}^{n}A_{ik} &\text{if $i=j$}\\
-A_{ij} &\text{if $i\neq j$}
\end{cases}
\end{equation*}
 
\par In the case of $\mathcal{L}_{in}$ now, its diagonal entries contain the in-degree strengths of the nodes. The solution to (7) is: 
 
  \begin{equation}
\label{Eq:equ8}  
 \mathbf{x}\left(t\right)=\ e^{{-L}_{in}t}\mathbf{x}\left(0\right)
  \end{equation}   
 
\par For balanced networks, where the sum of the weights of in-degrees are equal to the sum of the weights of out-degrees for all nodes, the advection and consensus dynamics are equal, i.e. Eq (5) is equal to Eq (8) and both are equal to heat diffusion. For the purposes of our adaptive rewiring algorithm, we use the exponential part of (5) for advection, which indicates the dynamics when we project unit output to a node (while the rest are zero) for all the nodes in parallel. We call this part $\alpha(t)$, the advection kernel: 
  
\begin{equation}
\label{Eq:equ9}   
\alpha\left(t\right)=\ e^{{-L}_{out}t}I_{nxn}=e^{{-L}_{out}t}  
 \end{equation}   
  
Similarly, the consensus kernel, $c(t)$, is:  

\begin{equation}
\label{Eq:equ10} 
c\left(t\right)=\ e^{{-L}_{in}t}I_{nxn}=e^{{-L}_{in}t}  
 \end{equation}

\subsection{Adaptive rewiring algorithm}

Before the onset of the adaptive rewiring algorithm, the initial network is $D\ =\ D_{random}$, a network with $N = 100$ nodes and $m=\left[2\log{\left(n\right)}\left(n-1\right)\right]=\ 912$ randomly assigned to pairs of nodes, with the only exception being that the node cannot point to itself. The network size and the number of connections have been tested for undirected networks in previous studies \cite{Rentzeperis} \cite{Jarman}. Adaptive rewiring proceeds as follows for in-degree connections (and analogously for out-degrees):
\newline
\textbf{Step 1.}Select with uniform probability a node k from the nodes with nonzero, but not n-1, in- and out-degrees.
\newline

\textbf{Step 2.}Delete edge (i\textsubscript{2}, k) and add edge (i\textsubscript{1}, k). With probability p\textsubscript{random} select i\textsubscript{1} and i\textsubscript{2} based on the criteria of step 2.1 (random rewiring) otherwise select them based on step 2.2 (instructed rewiring).
\newline
\textbf{Step 2.1.} i\textsubscript{1} is selected randomly from the set (i,k)\( \notin \)E, i.e. nodes that are not in the in-degree neighborhood of k.  i\textsubscript{2} is selected randomly from the set (i,k)\(\in\)E, i.e. nodes that are in the in-degree neighborhood of k.
\newline
\textbf{Step 2.2.} Calculate the kernel of the algorithm used, f(t). From the set of nodes (i,k))\( \notin \)E, i\textsubscript{1} is the one with the highest concentration transfer with k. From the set (i,k)\(\in\)E, i\textsubscript{2} is the one with the lowest concentration transfer with k. Mathematically, this is expressed as follows (f($\tau$) function represents either  $\alpha$($\tau$) or c($ \tau$)):

 \begin{equation}
\label{Eq:equ11}
 i_{1} =  argmax_{(i,k)\notin E,     i \neq k }f_{ik}(\tau)
  \end{equation}
 \begin{equation}
\label{Eq:equ12}
 i_{2} =  argmin_{(i,k)\in E,     i \neq k }f_{ik}(\tau)
 \end{equation}

\textbf{Step 3.}  Go back to step 1 until r edge rewirings have been reached.
\newline
\par We refer to the time variable of the kernel as the rewiring rate ($\tau$), since before each rewiring we let the concentrations spread for $t = \tau$. We found that for different rewiring rates, the rewired networks converge to similar topologies. Thus, in the Results section we show rewired networks for fixed $\tau = 1$. Unless otherwise stated, for each rewiring run we perform r = 4000 rewirings.

\subsection{Measurement of path length}

The path length, L\textsubscript{ij} defined as the distance between node pairs i and j along their shortest paths. The average path length of a network is defined as:

 \begin{equation}
\label{Eq:equ13}
L_{ave}=\ \sum_{i\neq j}\ L_{ij}
  \end{equation}

\par The drawback with this definition is that we cannot include node pairs that are not connected via a path between them, for a pair of that sort has infinite path length. To remedy this, we first measure the efficiency of the network. Efficiency is defined as the inverse of path length: 

 \begin{equation}
\label{Eq:equ14}
E_{ij}=\frac{1}{L_{ij}}\ 
  \end{equation}

\par For this measure, an ordered pair of nodes yields zero efficiency if there is no path in the network from the first node in the pair to the second. Following this definition for the efficiency of a single pair of nodes, the average efficiency of the whole network -and for n possible node pairs-is then 

 \begin{equation}
\label{Eq:equ15}
E_{ave}=\ \frac{1}{n}\sum_{i\neq j}\ E_{ij}
  \end{equation}

\par After measuring E\textsubscript{ave}, we take its inverse to estimate a path length metric. 

\section{Results}
\subsection{Advection and consensus create winner-take-all hubs}

We identify a set of nodes as a convergent-divergent unit, if they can receive input from a pool of nodes in the network and provide an output back to the pool (Fig. 2). We first investigated which parts of convergent-divergent units arise from iteratively rewiring an initially randomly connected network, using solely advection or consensus as a basis for rewiring.  For advection-based rewiring, a characteristic topology evolved when candidate nodes’ out-degree connections were rewired; the resulting network contained a small subset of convergent hub nodes, which received connections from all other nodes (Fig S1A). Conversely, random topology was preserved when advection-based rewiring was applied to the in-degree neighborhoods (Fig S1B). For consensus-based rewiring of the in-degree neighborhoods of the nodes, this time the resulting network obtained a characteristic structure where a small subset of the nodes, acting as divergent hubs, projected their connections to all other nodes (Fig S1D).  Rewiring the out-degree neighborhoods based on consensus, by contrast, preserved the network’s random topology (Fig S1C).

\begin{figure}
\centering
\includegraphics[]{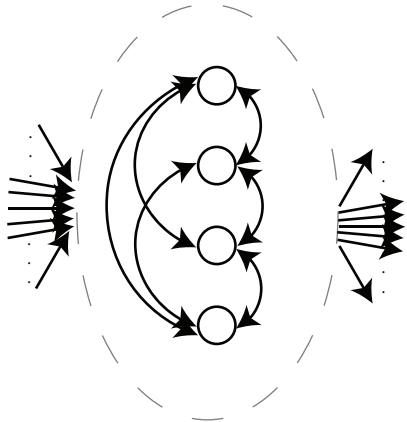}
\caption{A schematic representation of a convergent-divergent structure. A subset of the nodes receives all the nodes’ input, process it, and outputs it back to the nodes.}
\label{a2:fig:Res}       
\vspace{-0.4cm}
\end{figure}

\par Advection and consensus rewiring affords networks with convergent and divergent hubs respectively, the main building blocks of convergent-divergent units.  This process is remarkably fast; when out-degree neighbours were rewired based on advection only, or in-degree neighbours based on consensus only, the network produced winner-take-all configurations already after 200 rewirings (Fig 3), a much faster convergence compared to the emergence of modular or centralized structures in undirected networks1,2. However, neither advection nor consensus by themselves produced convergent-divergent units. We thus ask: Can their combination produce convergent-divergent units?

\begin{figure}
\centering
\includegraphics[width=\textwidth,keepaspectratio]{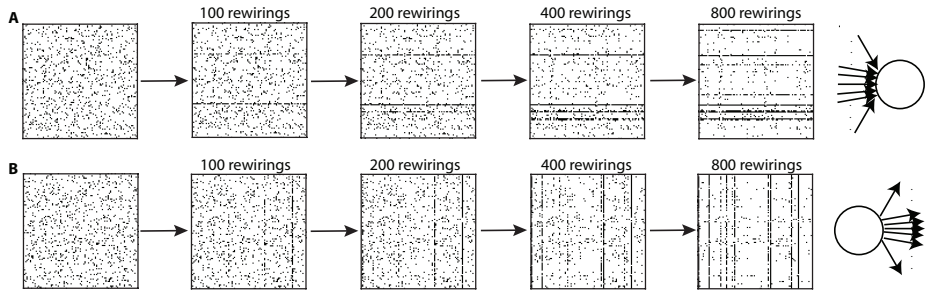}
\caption{Networks converge to winner-take-all configurations after few rewirings.
(A) The evolution of the adjacency matrix at different stages of rewiring when we use the advection algorithm and rewire the out-degree neighborhood. The rightmost illustration shows a schematic of a converging hub (B) Same as (A) but for the consensus algorithm when we rewire the in-degree neighborhood. The rightmost illustration shows a schematic of a diverging hub.}
\label{a3:fig:Res}       
\vspace{-0.4cm}
\end{figure}

\subsection{Combined advection and consensus effects improve node-to-node communication}

\par To study the combined effects of advection and consensus in one rewiring scheme, for each iteration of the rewiring algorithm we randomly chose with probability p(advection) to rewire based on advection the out-degree neighborhood of a candidate node, and with probability 1-p(advection) to rewire based on consensus the in-degree neighborhood of a candidate node. We found that as we increase p(advection), the network transitions from one dominated by divergent hubs to one dominated by convergent hubs: the vertical connectivity in the adjacency matrix transitions to a horizontal one (Fig S2). For any combination of advection and consensus, rewiring resulted in a sharp decrease in the path length and an increase in the number of nodes connected via a path compared to the networks rewired solely by advection or consensus (Fig S3). 
\newline
\par We probed further the mixed advection-consensus algorithm for the specific case of balanced rewiring, when rewiring was based with equal probabilities on either advection or consensus. This scheme offers a level playing field, on which networks can develop a balance of hub structures with in-coming or out-going connections. Maintaining the balance of advection and consensus, we introduced various proportions P\textsubscript{random} of random rewirings to the algorithm. That is, at each rewiring we randomly chose with 1- P\textsubscript{random} a rewiring based with equal proportions on either advection or consensus or with P\textsubscript{random} a random rewiring. Previous findings have shown that regular network topologies with some random connections can, to a large extent, maintain their structural properties while also becoming more efficient in node-to-node communication \cite{Watts}. In line with that, we found that random rewiring further decreases the average path length (Fig S4A) and increases the number of possible paths (Fig S4B) in the emerged networks. In all cases, we define the average path length as the inverse of the average efficiency of the network. This enabled us to incorporate in the measure the node pairs that do not have a path between them (see Methods subsection: Measurement of Path Length). We found that the average path length of all possible node combinations (cyan line in Fig 4) decreases precipitately as P\textsubscript{random} increases (1/\textsubscript{Eave,All} = 5.28, 4.66, 3.15, 2.44, 2.17 for P\textsubscript{random} = 0, 0.2, 0.4, 0.6, 0.8 respectively).  The average paths lengths between nodes that are connected by a path in the network (blue line in Fig 4) do not change to a large extent (1/\textsubscript{Eave,Paths} = 2.42, 2.71,2.37, 2.17 for P\textsubscript{random} = 0.2,0.4,0.6,0.8 respectively). Thus, the decrease can be attributed to a decrease in the proportion of unconnected node pairs (red bar in Fig 4).
\newline

\begin{figure}
\centering
\includegraphics[width=\textwidth,keepaspectratio]{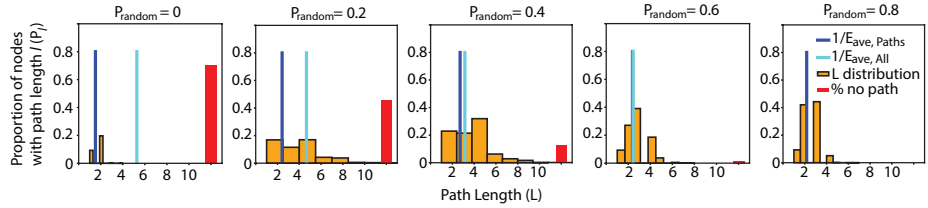}
\caption{The number of non-connected nodes decreases (available paths increase) as P\textsubscript{random} increases. 
Metrics for rewired networks based on balanced mixed consensus-advection algorithm for different P\textsubscript{random} values. The orange bars show the distribution of path lengths, the red bar the proportion of node pairs that do not have a path, the blue line shows the inverse of the average efficiency of the networks when considering only the node pairs with a path between them, and the cyan all possible node pairs. We consider the last two measures as path length metrics.}
\label{a4:fig:Res}       
\vspace{-0.4cm}
\end{figure}

\par We may consider as optimal those networks that have small path lengths (a topology that is essential for better node-to-node communication) and that still exhibit the structural properties effectuated by advection and consensus (convergent and divergent hubs). We find however, that as we increase the proportion of random rewiring and reduce the path length, we also reduce the number of hubs in the network (Fig 5A). 
\newline

\par To quantify the effects of those two opposing forces, we devised two measures we call structure efficiency metrics. High values correspond to networks with high efficiency in terms of node-to-node communication (small path length and large number of node pairs that are connected via a path) that also maintain to a large extent their structure (hub nodes). The first metric takes path length into account. This metric is defined as the ratio of the number of hubs over the path length. The second metric takes into account the number of connected nodes. This metric is defined as the number of connected nodes times the number of hubs. We find that the structure efficiency metric based on path length reaches a global maximum in absence of random rewiring, but it also has a local maximum for intermediate proportions of random rewiring (Fig 5B). When measuring structure efficiency based on available paths, we obtain a different pattern: the maximum for hubs with smaller in- and out-degrees (50, 60) is found at intermediate random rewiring proportions (Fig 5C). 
\newline

\begin{figure}
\centering
\includegraphics[width=\textwidth,keepaspectratio]{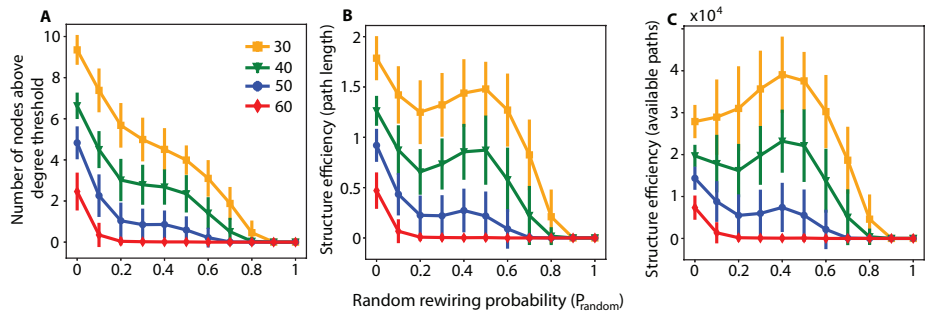}
\caption{Random rewiring can increase structure efficiency in terms of nodes’ connectivity (available paths).
(A) Number of hubs as a function of random rewiring. Hubs are defined as nodes with a minimum number of out-degrees (from 50 to 90) (B) Structure efficiency metric that uses path length as a function of random rewiring (C) Structure efficiency metric that uses the number of connected nodes (or available paths) as a function of random rewiring.}
\label{a5:fig:Res}       
\vspace{-0.4cm}
\end{figure}

\subsection{Convergent-divergent structures within the network}

\par Synthesizing our results so far, we are now able to obtain structures within the network that can be identified as convergent-divergent units. Convergent-divergent units receive input from local nodes, process it, and then outputs it back to the local nodes (Fig 2). Candidate hub nodes that are part of the convergent-divergent unit are the ones with a large number of degrees (both in- and out-degrees). We selected as hub nodes the ones with a minimum number of total degrees (the addition of in- and out-degrees had to be greater than a preselected value), that also had a greater than one in- or out-degree. Nodes outside of the convergent-divergent units were referred to as local nodes.
\newline
\par We require a path for local nodes to access the convergent-divergent unit, and for the hub nodes in the convergent-divergent unit to feed back to the local nodes. To this end, we probed the amount of the incoming flow from the local nodes to the convergent-divergent unit, and of the outgoing flow from the convergent-divergent unit back to the local nodes. For the incoming flow, we measured the proportion of local nodes with a path to at least one of the hub nodes. For the outgoing flow, we measured the proportion of local nodes that can receive input from the hub nodes. We found that the proportion of nodes with an incoming path to the convergent-divergent unit and of outgoing paths from the unit back to the nodes grows with increasing P\textsubscript{random}. The number of nodes that qualify to be in the convergent-divergent unit also grows with increasing P\textsubscript{random}, with P\textsubscript{random} $\geq$ 0.4 the proportion for ingoing and outgoing paths being greater that 0.9 (Fig 6A). 
\newline
\par We are also interested in the number of nodes that qualify as hub nodes and how random rewiring affects this number. We found that the number of hub nodes increases with increasing P\textsubscript{random} (Fig 6B). This runs counter to intuition since, in our previous analysis, we showed that random rewiring is detrimental to the build-up of hub structures. However, in our selection of hub nodes one requirement is that they have in- and out-degrees that are at least greater than 1. In absence of random rewiring, there are many nodes that either have only in-coming or out-going connections. This imbalance is counteracted by random rewiring. Removing the constraint that hub nodes need to have both in- and out-connections would reverse the trend, as the number of hub nodes then decreases with the proportion of random rewiring.
\newline
\par We subsequently examined the connections within the convergent-divergent unit that are in-between the hub nodes. We consider those that are part of the shortest path from a pair of converging to diverging hubs. For these intermediate units, for processing and transformation of incoming information from the local nodes, dense connectivity and ease of communication within the convergent-divergent substructure are desirable. We found that the connectivity density of the in-between nodes, we collectively call the intermediate unit, is greater than that of connected subnetworks of local nodes (same number of nodes for both subnetworks; P<0.001) randomly chosen from the same network (Fig S5A). This feature facilitates communication and computation within convergent-divergent units. Along with the hub nodes, which by definition form a dense subnetwork, the intermediate unit forms the connective core\cite{Shanahan} of the rewired network.  Note that the intermediate unit grows as P\textsubscript{random} increases (Fig S5B). 
\newline
\begin{figure}
\centering
\includegraphics[width=\textwidth,keepaspectratio]{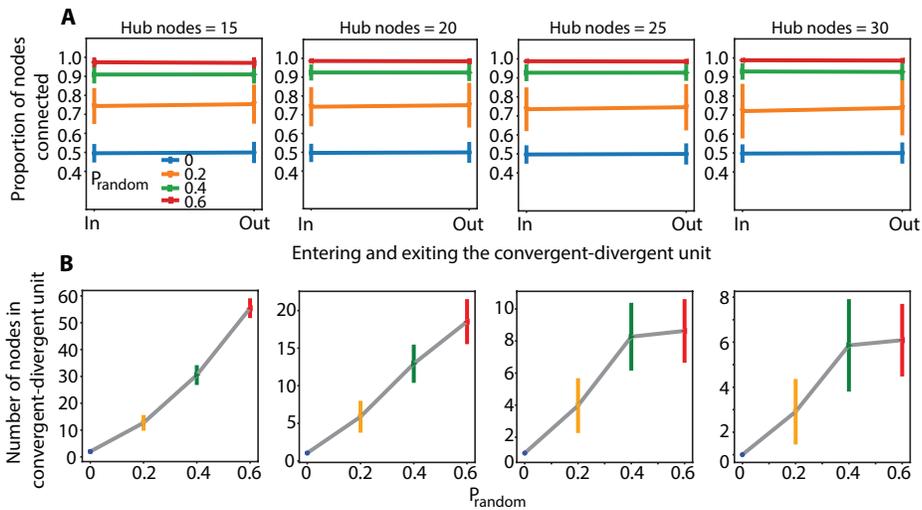}
\caption{Paths incoming to the convergent-divergent unit and out of the unit back to the nodes, as well as the number of nodes qualifying as part of the unit grow with increasing P\textsubscript{random}. 
A. Proportion of nodes with a path to the convergent-divergent unit, and nodes in the unit with an outgoing path back to the nodes. Left side (In) corresponds to the nodes with incoming connections to the convergent-divergent unit, and right side (Out) to the unit’s outgoing connections to the nodes. For the ‘In’ connections, a node is considered to be connected to the unit if one of its outgoing connections has a path to at least one of the hub nodes constituting the unit. For the ‘Out’ connections, a node is considered to be connected to the unit if at least one of the hub nodes in the unit has an outgoing path to it. A hub node and part of the unit is considered a node with a minimum number of in-going plus out-going connections. This number varies from left to right.
B. Average number of nodes in the unit as a function of P\textsubscript{random} for different hub nodes.} 
\label{a6:fig:Res}       
\vspace{-0.4cm}
\end{figure}     

\section{Discussion}
\par Convergent-divergent units are neural network core structures that receive their input from the local nodes and sends it back to them after processing. We investigated the emergence of convergent-divergent units within neural networks through adaptive rewiring based on spontaneous activity.  We found that adaptively rewiring of initially random directed networks leads to the emergence of converging-diverging units, when the flow of activity is represented by the directed equivalents of diffusion, advection and consensus. An additional proportion of random connections improves extensively node communication by decreasing the average path length and increasing the number of available paths while maintaining the converging-diverging structure. 
\newline
\par For adaptive rewiring to produce converging-diverging units in neural network models, the application of both advection and consensus is necessary. Advection-based rewiring of the out-degree neighborhoods produces convergent hubs; consensus-based rewiring of the in-degree neighborhoods of the nodes produces divergent hubs. The prominence of either type of hubs in the network depends on the relative frequency with which consensus or advection-based rewiring is chosen.   Convergent hubs are suitable for information integration, divergent ones for broadcasting. The proportion may therefore be varied depending on the position of a hubs in the information processing hierarchy (e.g. more divergence early in the perceptual processing stream: more convergence in higher-order areas).
\newline
\par We investigated balanced networks, networks with equal proportions of advection and consensus-based rewiring. The shortest path between each input (converging) and output (diverging) hub, could consist of a number of nodes that are not hubs; we call those intermediate nodes. We found that the intermediate nodes have a connectivity density that is higher than the local nodes outside the convergent divergent units. The greater density is a product of the adaptive rewiring, which adds connections to neighborhoods with intensive traffic flow. Their position between an input and output hub assures that intermediate nodes have an intensive in and outflow. The resulting dense connectivity serves the function of the networks of intermediate nodes as information processing cores.
\newline
\par Convergent-divergent units with abundant connectivity have been shown to offer a robust representation and propagation of input in a cascading structure \cite{Shaw}.  Moreover, convergence of variable inputs are pervasive in the brain, for example neurons in superior temporal sulcus receiving a multimodal sensory input \cite{Barraclough}. 
\newline
\par The present convergent-divergent structure, which feeds back to its input nodes is a mechanism that can also give rise to selectivities of sensory neurons that respond not only to local features but are also sensitive to long range contextual information. In visual processing, at least as early as in V1, neurons have been shown to respond to a number of stimuli outside of their local classical receptive field1, a property that is reflected in perception \cite{Rentzeperis3}. 
\newline
\par Previous applications of the adaptive rewiring principle on undirected networks showed topological effects emerging from aggregate activity that mirrored brain anatomy \cite{Rentzeperis}-\cite{Berg}\cite{Gong2}-\cite{Papadopoulos}. Namely, adaptive rewiring was sufficient for gradually transforming random undirected networks into structured architectures with features pervasive in the brain such as small worldness \cite{Bassett}\cite{Sporns}, modular connectivity \cite{Hilgetag}\cite{Bullmore}, and rich-club organization \cite{Heuvel}\cite{Zamora}. We studied directional networks, as they offer a more realistic representation of signal transfer \cite{Sheng}-\cite{Harlow}. Directionality of connections is important for establishing a processing hierarchy at the level of motifs \cite{Milo}\cite{White}, circuits \cite{Somers}\cite{Adesnik}, layers \cite{Mountcastle}, clustered axonal branches \cite{Leong}\cite{Oh}, and functional regions \cite{Parent}-\cite{Mogenson}, e.g. for processing sensory information \cite{VanEssen}\cite{Felleman} and for sensorimotor control \cite{Loeb}.  In addition, as shown here, networks of directed connections spontaneously evolve a core structure of convergent-divergent units under adaptive rewiring. Adaptive rewiring can therefore be considered a key principle of self-organization in brain networks.

\newpage
\begin{center}
\textbf{\large Supplemental Materials: Adaptive rewiring evolves brain-like structure in directed networks}
\end{center}

\setcounter{figure}{0}

\makeatletter 
\renewcommand{\thefigure}{S\@arabic\c@figure}
\makeatother

\begin{figure}
\centering
\includegraphics[width=\textwidth,keepaspectratio]{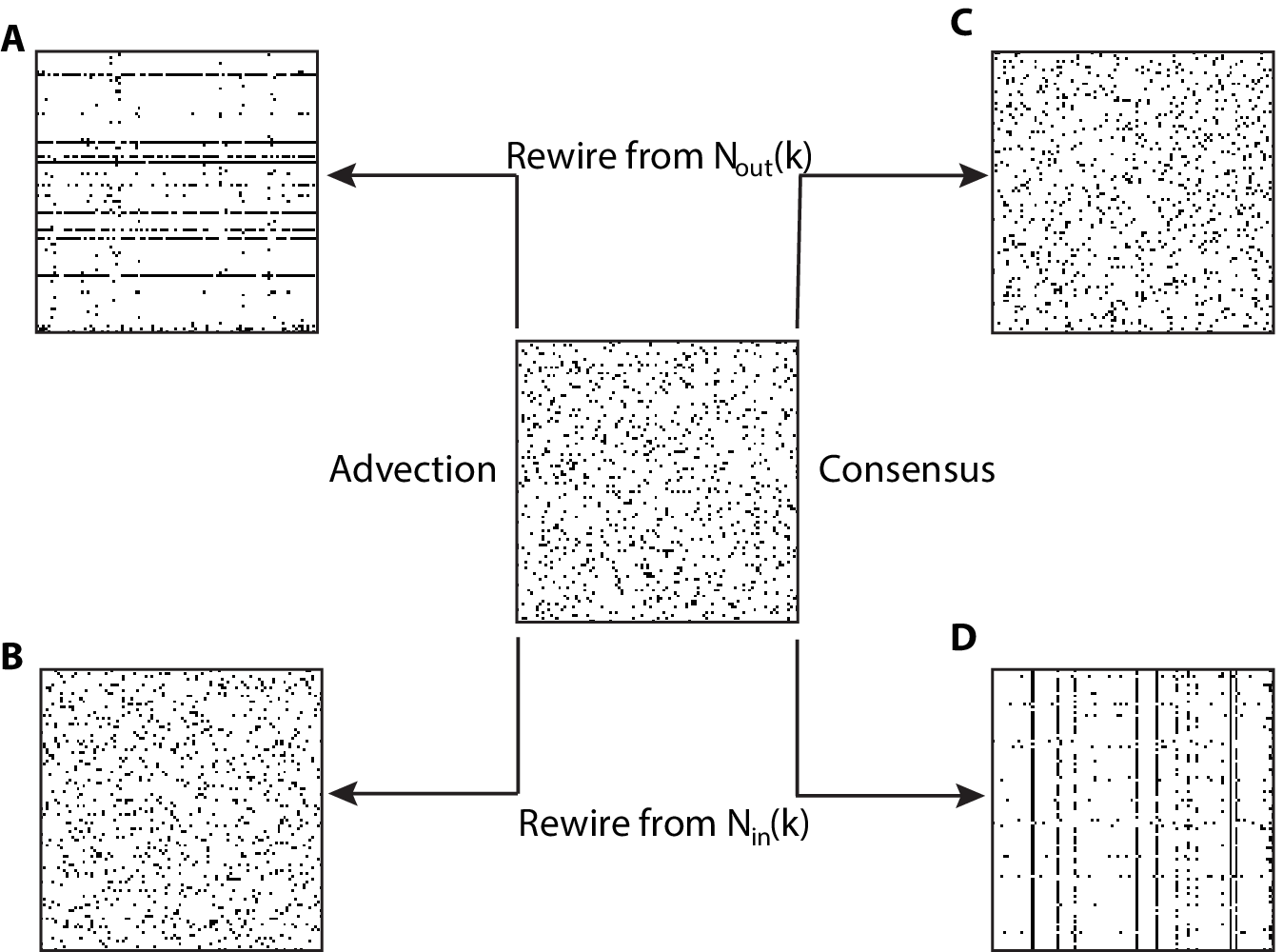}
\caption{Examples on the effects of advection (left side) and consensus (right side) on the evolution of a random network (adjacency matrix in the middle) when only the out-degree neighbors of the candidate nodes are rewired (top) or the in-degree neighbors (bottom).}
\label{s1:fig:Res}       
\vspace{-0.4cm}
\end{figure}

\pagebreak
\begin{figure}
\centering
\includegraphics[width=\textwidth,keepaspectratio]{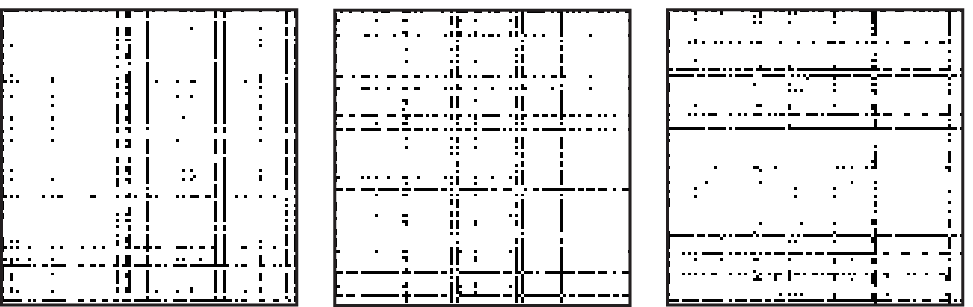}
\caption{As we increase p(advection), the vertical (out-degree hubs) connectivity transitions to a horizontal (in-degree hubs) one. From left to right: example adjacency matrices for p(advection) = 0.2, 0.5, 0.8.}
\label{s2:fig:Res}       
\vspace{-0.4cm}
\end{figure}

\pagebreak
\begin{figure}
\centering
\includegraphics[width=\textwidth,keepaspectratio]{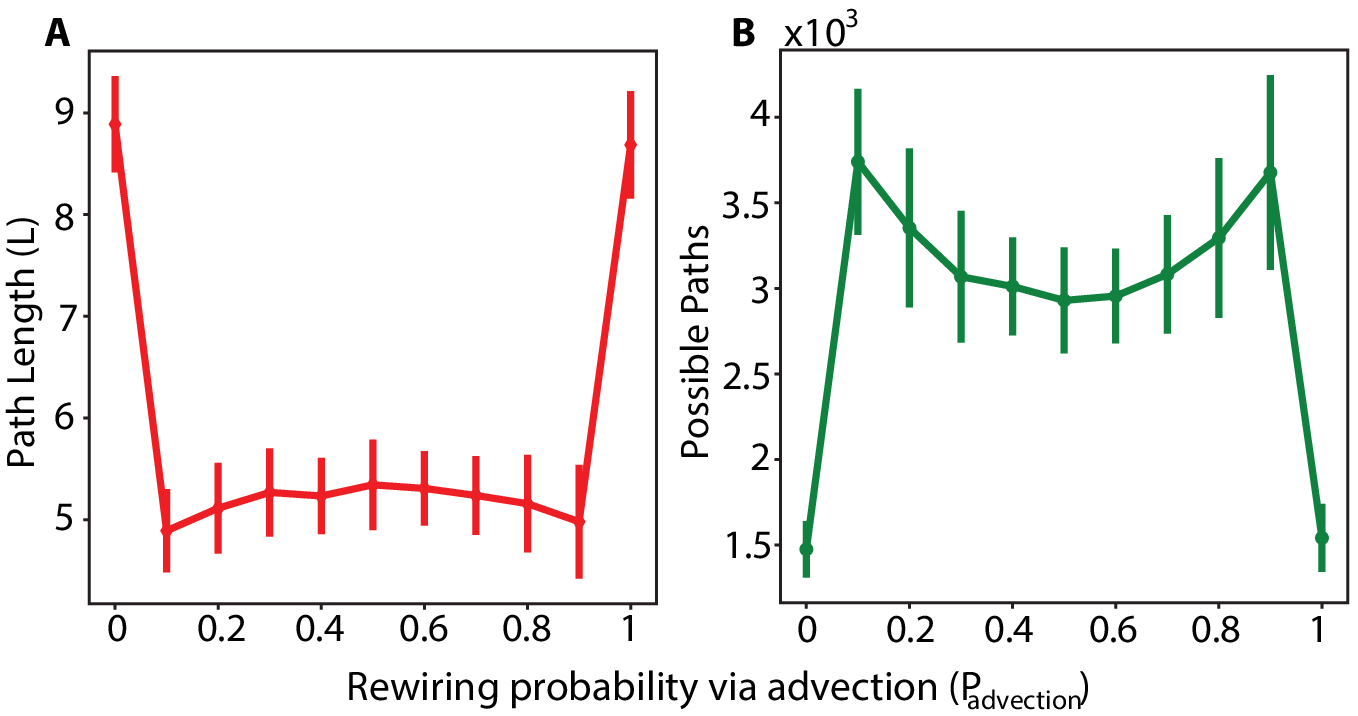}
\caption{The combination of advection and consensus in a rewiring scheme decreases path length and increases possible paths compared to rewiring based only on advection or consensus
(A) Average path length as a function of p(advection) (B) Possible paths as a function of p(advection). The data points and vertical lines in these and subsequent figures denote the means and standard deviations from 100 runs.}
\label{s3:fig:Res}       
\vspace{-0.4cm}
\end{figure}

\pagebreak
\begin{figure}
\centering
\includegraphics[width=\textwidth,keepaspectratio]{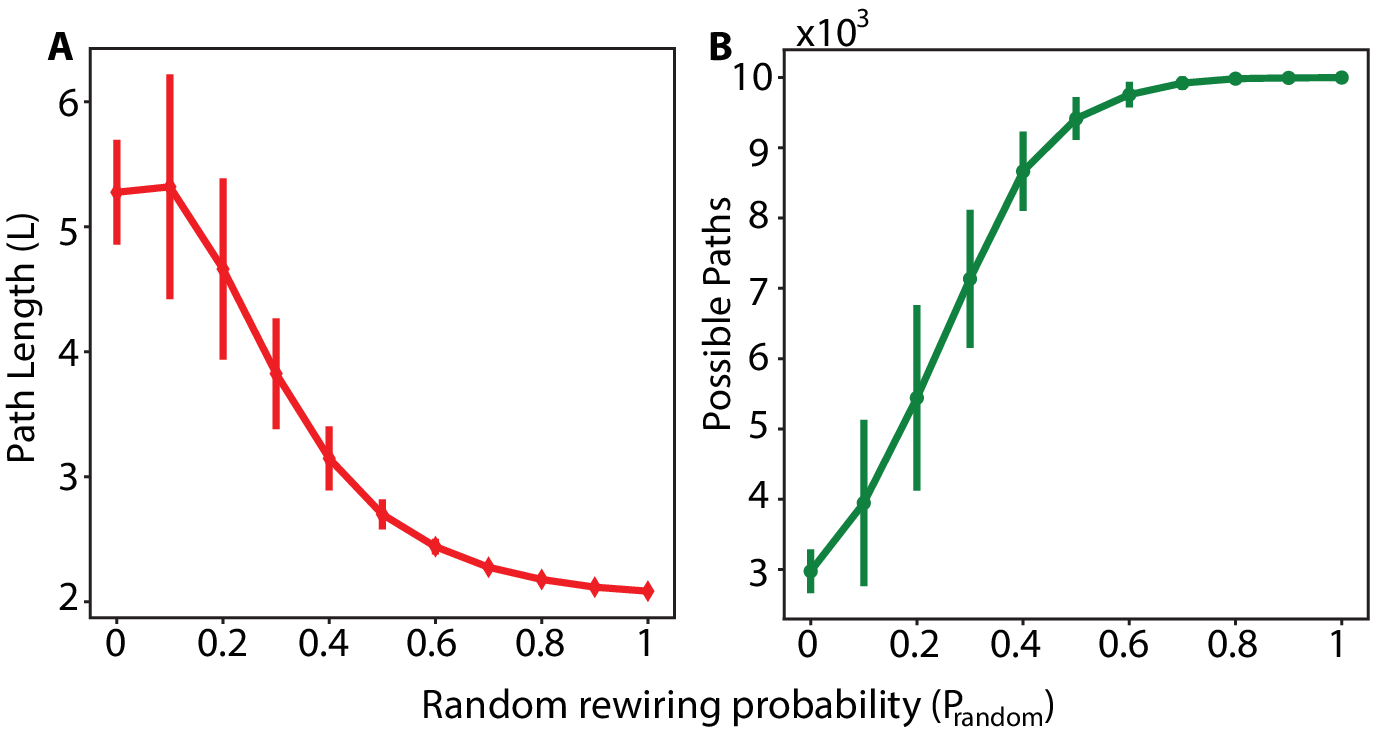}
\caption{In a combined advection/consensus scheme, random rewiring contributes in decreasing path length and increasing connectivity between nodes
 (A) Average path length  and (B) Possible paths as a function of P\textsubscript{random}.}
\label{s4:fig:Res}       
\vspace{-0.4cm}
\end{figure}

\pagebreak
\begin{figure}
\centering
\includegraphics[width=\textwidth,keepaspectratio]{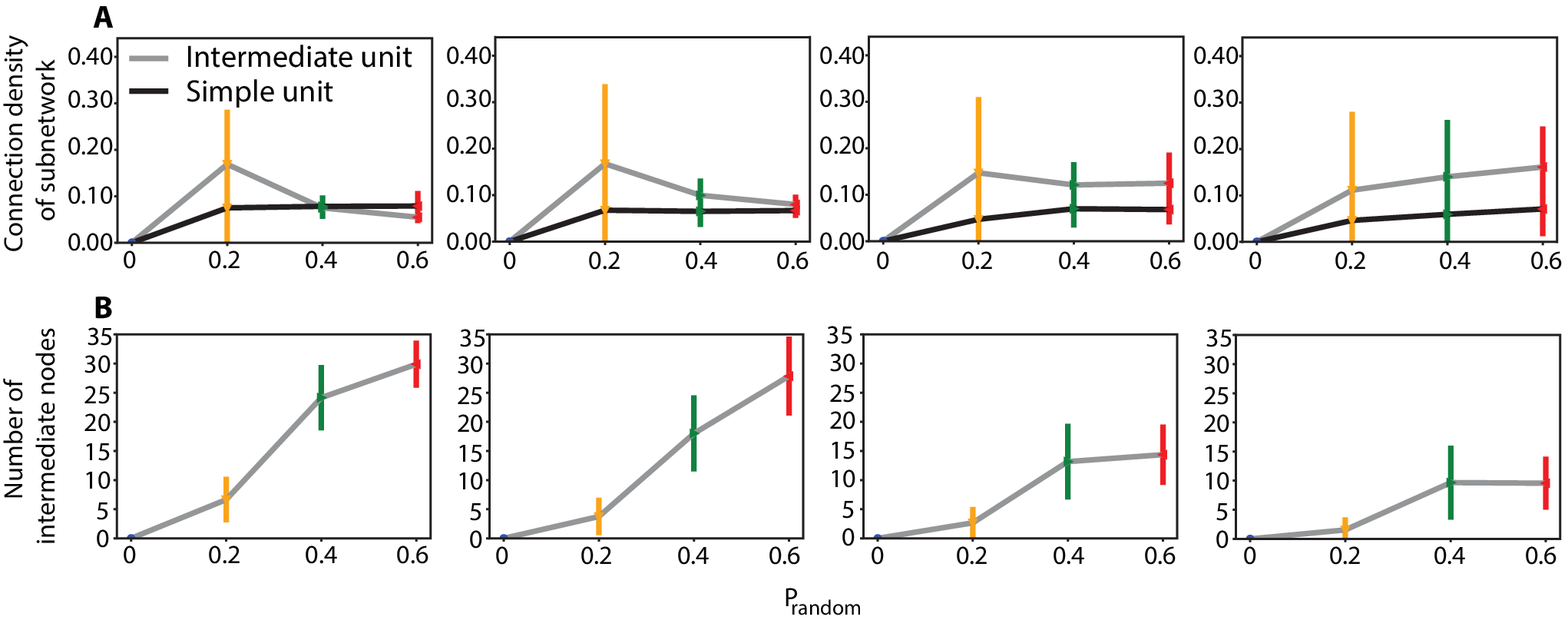}
\caption{The intermediate connections in a convergent-divergent unit are denser compared to a subset of simple nodes.
A. Density of the intermediate connections from the convergent-divergent unit and from a subset of simple nodes as a function of P\textsubscript{random} for different hub nodes criteria values. The nodes constituting the subset of simple nodes were selected randomly from the pool of all the simple nodes in the network. The number of nodes in the two substructures had to be equal.
B. Number of intermediate nodes as a function of P\textsubscript{random} for different hub nodes criteria values.}
\label{s5:fig:Res}       
\vspace{-0.4cm}
\end{figure}

\end{document}